\documentclass[12pt,letterpaper]{article}
\usepackage{float}
\usepackage{multirow, dcolumn}
\usepackage{ucs,amssymb,amsmath,mathtools}
\usepackage[utf8]{inputenc}
\usepackage{fontenc}
\usepackage{nicefrac}
\usepackage{graphicx}
\usepackage{changepage}
\usepackage[usenames,pdftex,dvips]{color,xcolor}
\usepackage[pdftex,colorlinks]{hyperref,ragged2e}
\usepackage[round, numbers, sort&compress]{natbib}

\usepackage[
singlelinecheck=false 
]{caption}

\newcommand{\va}[1]{\ensuremath{\mbox{Var}\left(#1\right)}}

\usepackage[margin=1in]{geometry}
\usepackage{setspace}
\usepackage[normalem]{ulem}

\newcommand{\Nb}{ \nobreak}

\DeclareMathOperator{\OR}{OR}
\DeclareMathOperator{\RR}{RR}
\DeclareMathOperator{\Var}{Var}
\DeclareMathOperator{\LLC}{LLC}
\DeclareMathOperator{\sech}{sech}
\DeclareMathOperator*{\argmin}{argmin}
\DeclareMathOperator\arctanh{arctanh}
\def\Asym{\stackrel{H_0}{\sim}}

\title{Put the odds on your side: a new measure for epidemiological associations}
\begin{document}
\maketitle
\thispagestyle{empty}
\begin{flushleft}
  \author{Olga A. Vsevolozhskaya, Dmitri V. Zaykin}\\
\end{flushleft}

\clearpage
\setcounter{page}{1}
\section*{ABSTRACT}

\noindent The odds ratio (OR) is a measure of effect size commonly used in observational research. OR reflects statistical association between a binary outcome, such as presence of a health condition, and a binary predictor, such as an exposure to a pollutant. Statistical inference and interval estimation for OR are often performed on the logarithmic scale, due to asymptotic convergence of log(OR) to normality. Here, we propose a new normalized measure of effect size, $\gamma'$, and derive its asymptotic distribution. We show that the new statistic, based on the $\gamma'$ distribution, is more powerful than the traditional one for testing the hypothesis $H_0$: log(OR)=0. The new normalized effect size is termed ``gamma prime'' in the spirit of $D'$, a normalized measure of genetic linkage disequilibrium, which ranges from -1 to 1 for a pair of genetic loci. The normalization constant for $\gamma'$ is based on the maximum range of the standardized effect size, for which we establish a peculiar connection to the Laplace Limit Constant. Furthermore, while standardized effects are of little value on their own, we propose a powerful application, in which standardized effects are employed as an intermediate step in an approximate, yet accurate posterior inference for raw effect size measures, such as log(OR) and $\gamma'$.
\vskip 2ex
\noindent \textbf{Keywords}: gamma prime coefficient; posterior interval estimation; standardized coefficient
\clearpage
\noindent Odds ratio (OR) is an ubiquitous measure of effect size in medical and epidemiological research. Among its useful properties is invariance across sampling designs (e.g., case-control or prospective), and straightforward interpretation in terms of logistic regression coefficients. Moreover, the log transformed odds ratios, log(OR), have analytically attractive properties, for example, the asymptotic distribution of log(OR) rapidly converges to a normal distribution as the sample size increases. Given the estimated log of odds ratio, $\log(\widehat{\OR})$, a common statistic used to assess significance of an association is:
\begin{eqnarray}
   Z &=& \frac{\log\left(\widehat{\OR}\right)} {\sqrt{\va{\log\left(\widehat{\OR}\right)}}} \Asym \text{Normal}(0,1)\\ 
 & = & \frac{\log\left(\widehat{\OR}\right)} {\sqrt{\sum{  \nicefrac{1}{n_{ij}}  }}},
 \label{Ztest}
\end{eqnarray} 
where $N=\sum n_{ij}$ is the sum of the four cell counts in a $2 \times 2$ table. The total sample size $N$ can be factored out and the statistic can be re-expressed as:
\begin{eqnarray*}
   Z &=& \sqrt{N} \, \, \frac{\log(\widehat{\OR})} {\hat{\sigma}}, \\
  \hat{\sigma} &=& \sqrt{\frac{1}{\hat{w}} \frac{1}{\hat{p}(1-\hat{p})} + \frac{1}{1-\hat{w}} \frac{1}{\hat{q}(1-\hat{q})}},
\end{eqnarray*}
where $\hat{w}$ is the sample proportion of cases, $\hat{p}$ is the estimated probability of exposure among cases, and $\hat{q}$ is the estimated probability of exposure among the controls. Thus, the asymptotically normal $Z$-statistic can be expressed as a product of the square root of the sample size ($\sqrt{N}$) and the standardized log(OR) ($\delta$) in the following way:
\begin{eqnarray*}
   Z &=& \sqrt{N} \times \hat{\delta}, \quad \hat{\delta} = \frac{\log(\widehat{\OR})} {\hat{\sigma}}. 
\end{eqnarray*}
In this paper, we derive the lower and the upper bounds on the possible values of the standardized effect size $\gamma=\delta_{\max}(\log(\OR))$ and show that $\gamma$ can not exceed the Laplace Limit Constant (LLC). Although this result was previously stated by our group \cite{vrz2017bayesian}, details and the derivation were never provided. Then, using the LLC as a normalizing constant, we propose a new measure of the effect size, $\gamma' = \nicefrac{\gamma}{\LLC}$, that is varying between -1 and 1. We derive an asymptotic distribution of $\gamma'$ and show, via simulation experiments, that the new association statistic based on the $\gamma'$ distribution is more powerful than the traditional one based on Eq. (\ref{Ztest}). Finally, we show how the standardized log(OR) can be utilized to accurately approximate posterior inference for the raw effect size, such as log(OR) and the newly proposed $\gamma'$.

\section*{METHODS}
\subsection*{Bounds for the standardized logarithm of odds ratio} \label{bounds}
We shall assume for now that epidemiological data is summarized by a 2$\times$2 table as:
\begin{table}[ht!]
\centering
\begin{tabular}{ccc}
\hline
&\multicolumn{2}{l}{Exposure status}\\
\cline{2-3} \rule{0pt}{3ex}
Disease status & $E$ & $\bar{E}$ \\
\hline \rule{0pt}{3ex}
$D$ & $n_{11}=n_D\hat{p}$ & $n_{12}=n_D(1-\hat{p})$\\
$\bar{D}$ & $n_{21}=n_{\bar{D}}\hat{q}$ & $n_{22}=n_{\bar{D}}(1-\hat{q})$\\
\hline
\end{tabular}
\label{tab1}
\end{table}

\noindent
where $n_{11}+n_{12} = n_D$ is the number of cases; $n_{21}+n_{22} = n_{\bar{D}}$ is the number of controls; and the number of exposed subjects is $n_{11}+n_{21}$. When sampling is random with respect to exposure, sample proportions $\hat{p}=\nicefrac{n_{11}}{n_D}$ and $\hat{q}=\nicefrac{n_{21}}{n_{\bar{D}}}$ are estimates of the population probabilities of exposure among cases and among controls, respectively, $p = \Pr(E|D)$ and $q = \Pr(E|\bar{D})$. Then, the effect of exposure on an outcome can be measured by the odds ratio, OR, defined as:
\begin{eqnarray*}
   \OR &=& \frac{ p/(1-p) }{ q/(1-q)  } \\
 &=& \frac{ \Pr\left(E \mid D\right) \left[1-\Pr\left(E \mid \bar{D}\right)\right]  }{\Pr\left(E \mid \bar{D}\right) \left[1-\Pr\left(E \mid D\right)\right]}.
\end{eqnarray*}
To study influence of various risk factors on the outcome, one can test the null hypothesis $H_0: \OR = 1$, or equivalently $H_0$: log(OR)=0. The logarithmic transformation is advantageous because of the bounded and asymmetric nature of OR (it can not take negative values) and also due to the fact that the distribution of log(OR) quickly converges to normality. Then, the classical test statistic is defined as: 
\begin{eqnarray*}
   Z & = & \frac{\log\left(\widehat{\OR}\right)} {\sqrt{\sum{  \nicefrac{1}{n_{ij}}  }}} \\ 
 & = & \sqrt{N} \, \, \frac{\log(\widehat{\OR})} {\hat{\sigma}}, \quad \text{where} \quad
  \hat{\sigma} = \sqrt{\frac{1}{\hat{w}} \frac{1}{\hat{p}(1-\hat{p})} + \frac{1}{1-\hat{w}} \frac{1}{\hat{q}(1-\hat{q})}},
\end{eqnarray*} 
and $\hat{w}$ is the sample proportion of cases, $\nicefrac{n_D}{N}$. 
The corresponding population parameter can be written as:
\begin{eqnarray}  
    \sigma^2 &=& \frac{1}{w} \frac{1}{p(1-p)} + \frac{1}{(1-w)} \frac{1}{q(1-q)}.  \label{eq:sd} 
 \end{eqnarray}
Conditionally on the value of OR, we can express variance ($\sigma^2$) as a function of two variables ($w$ and $p$) to emphasize that the standard deviation ($\sigma$) will vary depending on the study design and population prevalence of exposure among cases (we note that $q$ can be expressed in terms of $p$ and OR as $q = p/\left[ (1-p)\OR + p\right]$).
Alternatively, conditionally on the observed OR, one can express $\sigma$ in terms of the exposure probability,  $v=\Pr(E)$, and risk of disease among exposed as:
\begin{eqnarray}
  \sigma^2 &=& \frac{1}{v} \frac{1}{\Pr(D|E)\left[1-\Pr(D|E)\right]} \label{sigma.v} \\  \nonumber
  &+& \frac{1}{1-v} 
      \frac{1}{\Pr(D|\bar{E})\left[1-\Pr(D|\bar{E})\right]}, 
\end{eqnarray}
with $\Pr(D|\bar{E})=1/(1-\OR\left[1-1/\Pr(D|E)\right])$.

To obtain maximum possible value of the standardized $\log(\OR)$, we first need to minimize $\sigma$, conditional on the OR value, with respect to its two parameters. For example, if we set the first partial derivative of Eq. (\ref{sigma.v}) with respect to $v$ to zero, and solve the resulting equation in terms of $\Pr(D|E)$, it follows that:  
\begin{eqnarray}
  v_m &=& \argmin_v \sigma = \frac{1}{1 + \RR \,\, \sqrt{\OR^{-1}} }. \label{vm}
\end{eqnarray}
Further, setting the first partial derivative of Eq. (\ref{sigma.v}) with respect to $\Pr(D|E)$ to zero and plugging in $v_m$ instead of $v$ results in 
\begin{equation}
  \Pr(D|E)= 1 - \frac{1}{1 + \sqrt{\OR}}, \label{eq5}
\end{equation}
and
\begin{equation}
\Pr(D|\bar{E}) = \frac{1}{1 + \sqrt{\OR}} = 1 - \Pr(D|E). \label{eq6}
\end{equation}
Now, substituting Eqs (\ref{eq5}) and (\ref{eq6}) into Eq. (\ref{vm}), we obtain  $v = \nicefrac{1}{2}$. Similarly, operating with Eq. (\ref{eq:sd}), we can express the minimum $w$ value in terms of $p$ as
\begin{eqnarray}
   w_{m} &=& \argmin_w \sigma = \frac{1}{1 + \frac{p}{q} \,\, \sqrt{\OR^{-1}} }, \label{wm}
\end{eqnarray}
where $q = p/\left[ (1-p)\OR + p\right]$. Then, we can obtain an equivalent expression for $w$ as just we did for $v$, $w =  \nicefrac{1}{2}$.
Using the conditional value of $\sigma$, the maximum standardized log(OR) is
\begin{eqnarray}
   \gamma &=& \frac{ \log(\OR) }{ 2 \sqrt{2 + \frac{1+\OR}{\sqrt{\OR}}} }. \label{defgamma}
\end{eqnarray}
Using the identity $\frac{1+\OR}{\sqrt{\OR}} = 2 \cosh\left( \frac{\log(\OR)}{2}\right)$,
\begin{eqnarray}
   \gamma  &=& \frac{\log (\OR)}{2 \sqrt{2} \sqrt{1+\cosh \left(\frac{\log (\OR)}{2}\right)}} \\
 &=&  \frac{\log (\OR)}{4 \cosh \left(\frac{\log (\OR)}{4}\right)}.
\label{max.gamma}
\end{eqnarray}
Equation (\ref{max.gamma}) depends only on the logarithm of odds ratio, but it is not monotone in it: $\gamma$ reaches its maximum for log(OR) value at about 4.7987..., 
\begin{equation*}
\gamma_{\max} (\log(\OR) = 4.7987...) = 0.6627...
\end{equation*}
Surprisingly, as log(OR) exceeds that value, the corresponding normalized coefficient, $\gamma$, starts to decrease. Further, although Equation (\ref{max.gamma}) depends only on log(OR), its maximum can only be attained at the specific values of population parameters. Namely, (a) $v=w=  \nicefrac{1}{2}$, (b) $\Pr(D|\bar{E}) = 1 - \Pr(D|E)$ from Eq.(\ref{eq6}), which implies $\RR^2 = \OR$, and (c) $\log(\OR) = 4.7987...$, which implies log(OR) = 121.354$\dots$ and $\Pr(D|E) = \Pr(E|D) = 0.9167782798\dots$

\subsection*{Connection to the Laplace Limit Constant}
\begin{figure}[th!]
\centering
 \includegraphics[width=0.4\linewidth]{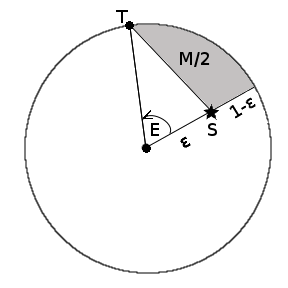} 
\caption{The Kepler equation: geometric interpretation. Given the knowledge of the area $M$ and the distance to the origin, $\varepsilon$, solve for the angle $E$ in $M = E - \varepsilon \sin(E)$.\cite{colwell1993solving}
}
\label{fig:KE}
\end{figure}

It turns out that there is an interesting connection between the expression for $\gamma_{\max}$ and the famous Kepler Equation (KE) for orbital mechanics, $M = E - \varepsilon \sin(E)$. Geometric interpretations of $M$, $E$ and $\varepsilon$ are illustrated by Figure \ref{fig:KE}. Specifically, suppose that one is inside a circular orbit, rescaled to be the unit circle, at the position S denoted by ``$\large{\star}$''. The shortest path to the orbit has length $1-\varepsilon$. A celestial body traveles the orbit from that point to point T. Given the area $M/2$ and distance $1-\varepsilon$, we want to determine the angle $E$. These three values are related to one another by Kepler's Equation. Planetary orbits are elliptical, so the actual orbit is along an ellipse inside of the unit circle.
Still, the calculation of the {\em eccentric anomaly}, $E$, is a crucial step in determining planet's coordinates along its elliptical orbit at various time points.

KE is transcendental, i.e., with no algebraic solution in terms of $M$ and $\varepsilon$, and it has been studied extensively since it is central to celestial mechanics. Colwell \cite{colwell1993solving} notes that ``in virtually every decade from 1650 to the present'' there have been papers devoted to the Kepler Equation in the book suitably named ``Solving Kepler's Equation over three centuries.'' The solution to KE 
involves the condition equivalent to Eq. (\ref{max.gamma}). Namely, the solution can be expressed as the power series in $\varepsilon$, provided $|\varepsilon \sin(E)| < |E-M|$ and that $\varepsilon < \psi / \cosh(\psi), \psi = |E-M|$, which is the ``Laplace Limit Constant,'' LLC \cite{plummer1918introductory}. The detailed mathematical derivation of the connection between Eq. (\ref{max.gamma}) and LLC is provided in ``Supplemental Materials (S-1).'' 

\subsection*{The proposed normalized measure of effect size and its distribution} \label{sec:gamma}
\begin{figure}[th!]
\centering
 \includegraphics[width=0.4\linewidth]{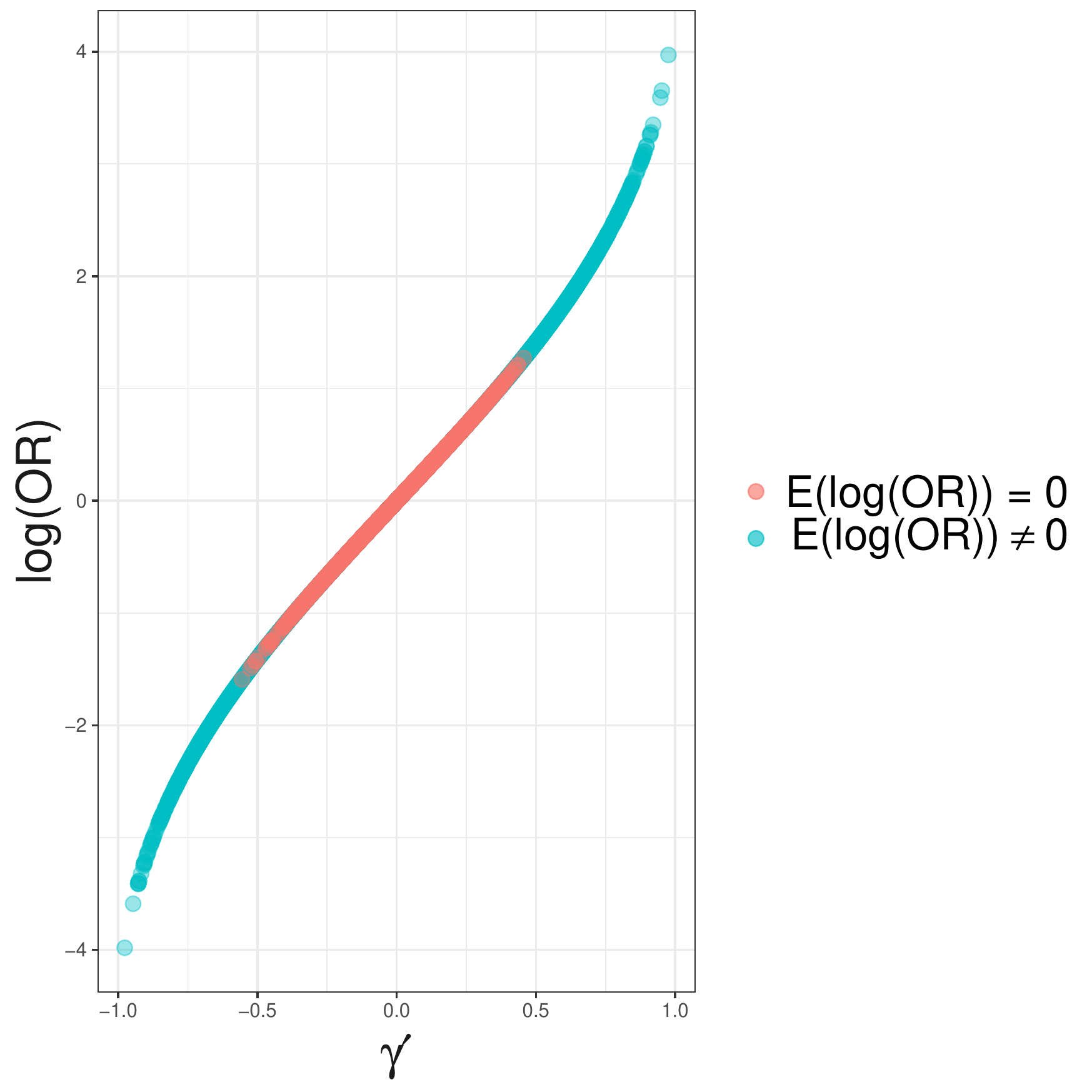} 
\caption{The relationship between log(OR) and $\gamma'$. The figure illustrates that under the null hypothesis (log(OR) $=$ 0), the relationship between sample values of log(OR) and $\gamma'$ is approximately linear, and under the alternative hypothesis (log(OR) $\neq$ 0) the relationship is monotone in the interval $\nicefrac{1}{121}\Nb <\Nb \OR\Nb <\Nb 121$.
}
\label{fig:logOR_gamma}
\end{figure}

As we showed above, at any value of $\log(\OR)$, the maximum of $\delta$ is
\begin{eqnarray*}
   \gamma = \frac{ \log(\OR) }{ 2 \sqrt{2 + (1+\OR) / \sqrt{\OR}} }.
\end{eqnarray*}
The bounded nature of $\gamma$ (ranging between $-$LLC and LLC) suggests a new normalized measure of effect size, $\gamma^{\prime} = \gamma / \LLC$, that has the range $-1\Nb \leq \Nb\gamma^{\prime}\Nb \leq \Nb1$. The new statistic is appropriate as a measure of effect size where it is monotone in $\OR$: within a very wide range of odds ratios: $\nicefrac{1}{121}\Nb <\Nb \OR\Nb <\Nb 121$. For instance, Figure \ref{fig:logOR_gamma} shows that under the null hypothesis, the relationship between log(OR) and $\gamma'$ is almost linear, and under the alternative hypothesis, the relationship is close to linear and monotone, as long as $\nicefrac{1}{121}\Nb <\Nb \OR\Nb <\Nb 121$ (these are rounded to integer OR values  before the LLC maximum is reached).

Although $\gamma^{\prime}$ is derived by using the range of the standardized $\log(\OR)$, it is not a standardized measure in the same sense as scaling by a standard deviation.  It is rather analogous to a coefficient denoted by $D^\prime$ \cite{lewontin1964interaction}, which is commonly used in genetics to measure association between alleles at a pair of genetic loci (linkage disequilibrium, LD). $D^\prime$ is akin to $\gamma^{\prime}$ in the sense that a raw measure of LD is divided by its maximum value (which is a function of allele frequencies) to yield the $-1 \le D^\prime \le 1$ range.

Using the first order Taylor series approximation, we derive an asymptotic variance of $\gamma^\prime$, as well as one- and two-sided asymptotic test statistics, as follows:
\begin{eqnarray*}
   \Var\left(\widehat{\gamma^\prime}\right) &=& \hat{\sigma}^2 \left(\frac{\sech\left[\nicefrac{\log\left(\widehat{\OR}\right)}{4}\right] 
\left(4 - \log\left(\widehat{\OR}\right) \tanh \left[\nicefrac{\log\left(\widehat{\OR}\right)}{4}\right] \right) }{ 16 \times \LLC } \right)^2. \\
  T &=&  \sqrt{N} \frac{\widehat{\gamma^\prime}}{\sqrt{\Var\left(\widehat{\gamma^\prime}\right)}},   \\
\end{eqnarray*}
which simplifies to
\begin{eqnarray}
T  &=& \sqrt{N}\frac{4 \log\left(\widehat{\OR}\right)}{ \hat{\sigma} \,\,\left(4 - \log\left(\widehat{\OR}\right) \tanh\left[ \nicefrac{\log\left(\widehat{\OR}\right)}{4} \right]\right)}.
\label{eq:test}
\end{eqnarray}
The asymptotic distributions for one- and two-sided statistics are
\begin{eqnarray*}
T &\Asym& \text{Normal}(0,1), \\
T^2 &\Asym& \chi^2_{(1)},
\end{eqnarray*}
where $\hat{\sigma}$ is defined as before:
\begin{eqnarray*}
   \hat{\sigma} &=& \sqrt{\frac{1}{\hat{w}} \frac{1}{\hat{p}(1-\hat{p})} +
     \frac{1}{1-\hat{w}} \frac{1}{\hat{q}(1-\hat{q})}}.
\end{eqnarray*}
We show by simulation experiments that the null distribution of this new statistic reaches the asymptotic chi-square quicker than the commonly used $X^2\Nb =\Nb \log(\widehat{\OR})^2\Nb /\Nb \hat{\sigma}^2$ and that the new statistic provides higher power under the alternative hypothesis.

We note that two other well-known transformations of the OR with the range from -1 to 1 are Yule's coefficients: $\mathcal{Y} = \frac{\sqrt{\OR}-1}{\sqrt{\OR}+1}$, the coefficient of colligation, and $\mathcal{Q} = \frac{\OR-1}{\OR+1}$ \cite{yule1912methods}.
Interestingly, using the identity $\frac{\sqrt{x}-1}{\sqrt{x}+1}=\tanh\left(\nicefrac{\log(x)}{4}\right)$, the statistic $T$ can be expressed as a function of $\mathcal{Y}$:
\begin{eqnarray}
   T  &=& \sqrt{N}\frac{4 \log\left(\widehat{\OR}\right)}{ \hat{\sigma} \,\,
      \left(\hat{\mathcal{Y}} - 4\right)}.
\label{eq:gamma_y}
\end{eqnarray} 
Further, note that $4 \arctanh\left(\mathcal{Y}\right) = 2 \arctanh\left(\mathcal{Q}\right) = \log(\OR)$. The $\arctanh$ transformation (to $\log(\OR))$, known as Fisher's variance stabilizing transformation \cite{fisher1915frequency}, is expected to improve the rate of asymptotic convergence to the normal distribution, thus we do not anticipate that the asymptotic test statistics based $\mathcal{Y}$ and $\mathcal{Q}$ would be competitive when compared to the $Z$ statistic based on the $\log(\OR)$. Nevertheless, we obtained approximate variances for $\mathcal{Y}$ and $\mathcal{Q}$ using the first order Taylor series approximation (the same type of approximation that yields the asymptotic variance for $\log(\OR)$) as follows:
\begin{eqnarray*}
   \widehat{\Var}\left(\widehat{\mathcal{Y}}\right) &=& \frac{1}{N}\left[\frac{\hat{p}}{\hat{w}(1-\hat{p}) \hat{q}^2 \left(\sqrt{\frac{\hat{p} (1-\hat{q})}{(1-\hat{p}) \hat{q}}}+1\right)^4}+\frac{1-\hat{q}}{(1-\hat{w}) (1-\hat{p})^2 \hat{q} \left(\sqrt{\frac{\hat{p} (1-\hat{q})}{(1-\hat{p}) \hat{q}}}+1\right)^4}\right], \\
   \widehat{\Var}\left(\widehat{\mathcal{Q}}\right) &=& \frac{1}{N}\left[\frac{4 (1-\hat{p}) \hat{p} (\hat{q}-1)^2 \hat{q}^2}{(1-\hat{w}) (\hat{p}+\hat{q}-2 \hat{p}\, \hat{q})^4}-\frac{4 (\hat{p}-1)^2 \hat{p}^2 (\hat{q}-1) \hat{q}}{\hat{w} (\hat{p}+\hat{q}-2 \hat{p}\, \hat{q})^4}\right].
\end{eqnarray*}
Via simulations, we confirmed that the statistic for $\mathcal{Y}$ tends to be more conservative and less powerful than $Z$, while the statistic for $\mathcal{Q}$ is anti-conservative and reaches the nominal 5\% size only around $N=1000$. However, these results are omitted here and we focus instead on comparisons of statistics based on $\gamma^{\prime}$ and $\log(\OR)$.

\subsection*{Approximate Bayesian inference}
The rationale for using standardized coefficients (e.g., standardized log odds ratio) as measures of effect size in epidemiologic studies has been questioned and it has been suggested that standardized coefficients are insufficient summaries of effect size \cite{greenland1986fallacy, greenland1991standardized}. Nonetheless, we argue that the standardized effects can be utilized efficiently in their new application developed here, as tools for delivering approximate Bayesian inference. Specifically, we propose to employ standardization as an intermediate step that yields posterior inference for parameters of interest (such as $\log(\OR)$ or $\gamma^\prime$). The key to this approach is the observation that it is often straightforward to obtain an approximate posterior distribution for standardized effects ($\delta=\nicefrac{\mu}{\sigma}$) using a noncentral density as likelihood. Once such standardized posterior distribution is estimated, it can be converted to an approximate posterior distribution for a parameter of interest, $\mu$.

Let $\xi = \sqrt{N}  \times \delta$ denote the noncentrality parameter of the raw effect size density (for instance, $Z \sim N(\xi, 1)$ or $X^2 \sim \chi^2_1(\xi)$). To obtain an approximate posterior distribution, one needs to specify a prior distribution for a raw measure of effect size, $\mu$, as a binned frequency histogram, with a finite mixture of values $\mu_1, \mu_2, \ldots, \mu_B$ (the mid-values of bins) and the corresponding probabilities, $\Pr(\mu_i)$ (the height of bins as percent values). For example, if the effect size is measured by $\mu$=log(OR), such binned frequency histogram may be bell-shaped with a sizable spike around zero, indicating that the majority of risk effects are anticipated to be small. Alternatively, if the effect size is measured by log$^2$(OR), the frequency histogram may be L-shaped, with a spike of the mass again at about zero. 

Next, we employ an approximation to a fully Bayesian analysis (which would have required a joint prior distribution for both $\mu$ and $\sigma$), and ``dress'' the raw parameter, by plugging in the estimate of the standard deviation, to obtain values of $\delta_i = \nicefrac{\mu_i}{\hat{\sigma}}$ and $\xi_i = \sqrt{N} \delta_i$. Then, given the observed value of a test statistic $T=t$, the posterior distribution of the standardized effect size will also be a finite mixture, calculated as:
\begin{eqnarray}
   \Pr(\xi_j \mid T = t) = \frac{\Pr(\mu_j) f(T = t \mid \xi_j)}{\sum_{i=1}^B \Pr(\mu_i) f(T = t \mid \xi_i)}, \label{postd}
\end{eqnarray}
where $f$ is the test statistic density with the non-centrality parameter $\xi_i, \; i = 1, \ldots, B$. Once the posterior distribution for the standardized effect size (times $\sqrt{N}$), is evaluated, one can approximate the posterior distribution for the raw parameter of interest by ``undressing'' it, i.e., multiplying by the sample standard deviation and scaling by the square root of the sample size. For example, 
\begin{eqnarray}
   \Pr(\gamma'_i \mid T=t) = \Pr\left(\xi_i \cdot \sqrt{\widehat{\Var}\left(\widehat{\gamma^\prime}\right)}  / \sqrt{N} \mid T=t\right)
\end{eqnarray}
or 
\begin{eqnarray}
   \Pr(\log(\OR)_i \mid T=t) = \Pr\left(\xi_i \cdot \sqrt{\widehat{\Var}\left(\log(\widehat{\OR})\right)} / \sqrt{N} \mid T=t\right) .
\end{eqnarray}
From this approximate posterior distribution, one can then obtain an effect size estimator as the posterior mean by taking a weighted sum (e.g., $E(\gamma' \mid T=t) = \sum_{i=1}^B \gamma'_i \Pr(\gamma'_i \mid T = t)$), construct posterior credible intervals, etc. `Approximation' here refers to approximating a fully Bayesian modeling: our approach is a compromise between the frequentist and the Bayesian methodologies due to the usage of plug-in frequentist estimates for certain parameters. Although the posterior distribution for the raw effect size is approximate (due to plugging in a point sample estimate of the standard deviation), it is nevertheless highly accurate, as we demonstrate through our simulation experiments. 

\section*{RESULTS}
\subsection*{Simulations: Frequentist properties} \label{sec:pwr}
To investigate statistical properties of the proposed procedures in relation to the traditional $Z$-test, we now turn to simulation experiments (the details of the simulation setup are provided in ``Supplemental Materials (S-2)''). 

Our simulations are not intended to be comprehensive, and we specifically compared only the two statistics discussed in this report. The comparison is `apples to apples,' i.e., between two similarly derived Wald test statistics, both based on transformations of OR as a measure of effect size. The basic model for OR, without stratification or covariates, follows from  $2 \times 2$ contingency table, and we note that performance of different tests for contingency table associations has been thoroughly investigated before \cite{larntz1978small,zaykin2008correlation}.

The Type-I error rates of the two tests, calculated under the null hypothesis of no effect, $\log(\OR)\Nb =\Nb 0$, are shown in Table \ref{type1}. For small number of cases, both statistical tests behave conservatively, but the size of the test based on $\gamma^\prime$ is considerably closer to the nominal level of $\alpha = 0.05$. As the number of cases increases, the size of both tests approaches the nominal level. Table \ref{pwr} shows statistical power of the two tests for the different combinations of $\log(\OR)$, its variance, and the number of cases. For all combinations of parameters considered, the $\gamma'$-based test has higher statistical power than the $Z$-test, particularly for small sample sizes. 

\begin{figure}[th!]
\centering
\includegraphics[width=0.4\linewidth]{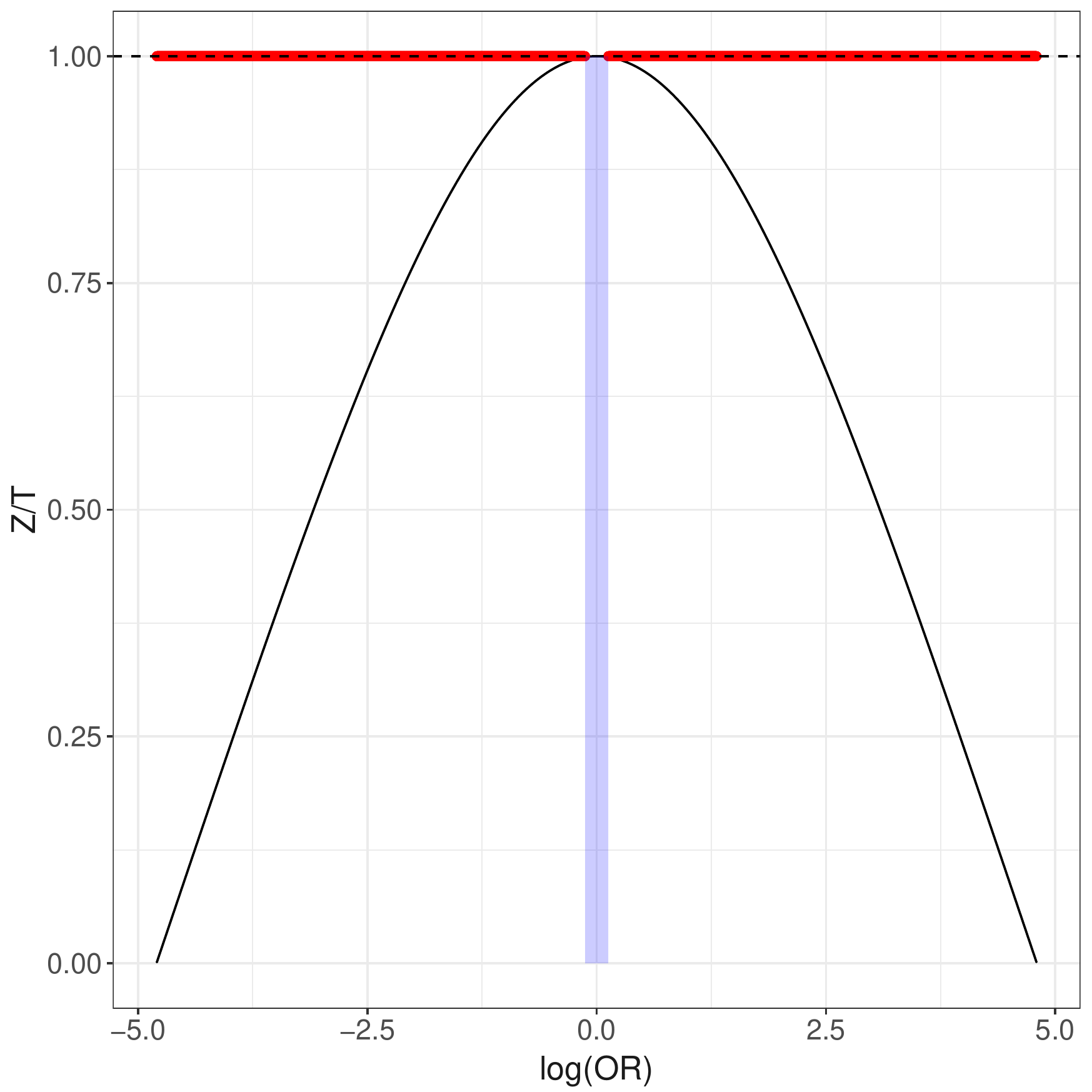} 
\caption{The range of $Z/T$-ratio values for $\nicefrac{1}{121}\Nb <\Nb \OR\Nb <\Nb 121$. The red line highlights log(OR) values, for which $\gamma'$-based $T$-statistic considerably exceeds $Z$-statistic. The blue rectangular highlights log(OR) values near the null hypothesis, for which the two statistics are similar to one another. Note that for all values of log(OR), $Z$-value never exceeds $T$-value.
}
\label{fig:r}
\end{figure}

We further note that the power of these two-sided tests can be investigated analytically by plugging in the population parameters, $p,q,w$ and considering the ratio of $Z$ and $T$ values, $Z/T$. Sample size and variance cancel out and their ratio becomes only a function of log(OR):
\begin{eqnarray*}
  Z/T = \frac{4 - \log(\OR) \tanh \left[ \nicefrac{\log\left(OR \right)}{4} \right] }{4}.
\end{eqnarray*}
Figure \ref{fig:r} illustrates that for all odds ratio values within the $(\nicefrac{1}{121} - 121)$ interval, $T$-statistics are at least as large as $Z$-statistics. Under the null (true log(OR) = 0), the ratio $Z/T$ is one, and the two statistics are equivalent. 

\subsection*{Simulations: Bayesian properties}
Averaged across simulations, Table \ref{tab:maxZ} reports (a) the true mean value of the raw parameter, $E(\gamma')$, corresponding to the maximum observed test statistic; (b) the posterior expectation  $E(\gamma^{\prime}\Nb\mid\Nb Z_{\max})$; (c) the average for the frequentist estimator of gamma prime, $E(\widehat{\gamma'})$; and (d) the average probability to contain the true log(OR) value by the high posterior density interval. 

On the one hand, Table  \ref{tab:maxZ} shows that posterior expectation is very close to the true average effect size value, even when posterior inference was performed using small sample sizes and extreme selection (i.e., the top-ranking result taken out of one million statistical tests). On the other hand, Table  \ref{tab:maxZ} illustrates that the frequentist estimator of $\gamma'$ is subject to the winner's curse and grossly exaggerates the true magnitude of effect size. Finally, after comparing posterior convergence to the corresponding nominal levels, it is clear that the posterior interval's performance is satisfactory. 

\subsection*{Real data application}
To illustrate a practical implementation of the proposed effect size measure, we calculated $\gamma'$ for six dietary intake risk factors associated with diabetes: whole grain \cite{fung2002whole}, protein \cite{tinker2011biomarker}, alcohol consumption \cite{cullmann2012alcohol}, fruits and berries \cite{montonen2005food}, dietary magnesium \cite{hruby2014higher}, and dietary calcium \cite{van2006dietary}. All reported associations used in this application reached nominal 5\% statistical significance. Reports that examinine individual dietary factors in relation to the type 2 diabetes (T2D) are inconclusive \cite{hu2001diet, liu2004prospective, song2004dietary} and to check robustness of these results, we converted the reported ORs to $\gamma'$'s and constructed posterior intervals for our new measure, assuming three different \textit{a priori} levels of belief that a reported dietary factor is truly a risk of T2D. The \textit{a priori} assumptions were set to (1) optimistic or 25\% chance that the reported association is false, (2) even or 50\% chance that the report is false, and (3) poor or 75\% chance that the report is false. For real associations, we assumed that there is 5\% {\em a priori} chance of encountering OR $\ge$ 2 (or OR $\le \nicefrac{1}{2}$). This is the same assumption as we made in our simulations (\ref{simulations}).

Table 4 summarizes our results. The first row of the table reports robustness of the whole grain intake association with T2D. The initially published frequentist estimate OR = 0.70 corresponds to the negative $\gamma'=-0.13$, indicating that greater whole grain intake may reduce risk of T2D (i.e., negative values of $\gamma'$ imply a protective effect). The posterior expectation of $\gamma'$ is similar to the frequentist estimate for all levels of \textit{a priori} belief that the effect is genuine. Furthermore, regardless of levels of uncertainty in the protective effect of the whole grain, the upper bound of the 95\% posterior interval did not cover 0, suggesting that the effect might be real or at least strong enough statistically to withstand substantial perturbations in prior assumptions.

The second row of Table 4 shows robustness of association between protein intake and T2D. The initial OR = 1.16 corresponds to the positive value of $\gamma' = 0.06$, suggesting that higher intake of dietary protein is associated with an increased risk of T2D. Posterior estimates for the dietary protein effect magnitude are not as high as the frequentist's for all levels of \textit{a priori} belief that the effect is genuine, but maintain posterior `significance' under the optimistic \textit{a priori} scenario (75\% prior chance that the association is real). For the 50-50 or lower \textit{a priori} chances, we can no longer conclude with confidence that dietary protein intake is associated with an elevated risk of T2D.

The remaining rows of Table 4 report results for alcohol, fruits and berries, dietary magnesium, and dietary calcium associations with T2D. At all levels of \textit{a priori} skepticism/optimism regrading the nature of the reported associations, the posterior intervals for $\gamma'$ values cover zero, indicating that these findings do not have a strong statistical support.

It is useful to contrast posterior intervals with the frequentist (confidence) intervals, CIs. None of the reported CIs, except for the dietary calcium association, cover zero. Can one modify prior assumptions so that posterior intervals would match CIs? In other words, what are the implicit prior assumptions that govern the width of CIs? It turns out that simply lowering the prior $\pi_0 = \Pr(H_0)$ from 25\% to 0\% is insufficient, and one also needs to assume a substantially flatter distribution for chances of a true association. Specifically, to match the endpoints of the reported CIs, one needs to assume a 5\% \textit{a priori} chance of encountering an OR as large as 4 for dietary risk factors of T2D and simultaneously lower $\Pr(H_0)$ to about 0.5\%. Although such prior assumptions are often described as non-informative or vague, they in fact correspond to a strong but unrealistic belief that observing large OR values is not much less likely than the small ones.

\section*{DISCUSSION}
In this article, we propose a new powerful transformation of the odds ratio to measure the magnitude of associations between risk factors and a binary outcome. The proposed test statistic for $\gamma'$ has competitive power compared to the traditional statistic for testing the null hypothesis that OR=1. Further, we introduce a simple and efficient approach for obtaining an approximate posterior distribution for $\gamma'$ and demonstrate its robustness to selection bias, a feature that aims to improve reliability of reported findings. Via simulations, we showed that $\gamma'$-based test has better control of the Type I error rate over the traditional $Z$-test for small sample sizes. The power of our method is always at least as good as that of the traditional $Z$-test under the tested scenarios. 

Our new measure $\gamma'$ is normalized by the Laplace Limit Constant to range between $-1$ and 1, yet it should not be regarded as an approximate transformation to a correlation coefficient, nor does it behave as a standardized measure of effect size. That said, the usual standardized log(OR) can be approximately related to the standardized slope and to the correlation coefficient  $R\Nb =\Nb \beta\Nb\times\Nb (\sigma_X\Nb /\Nb \sigma_Y)$ in simple linear regression models. When both $X$ and $Y$ are binary, $R$ can be expressed as:
\begin{eqnarray*}
   R &=& (p - q) \frac{ \sqrt{w(1-w)} }{ \sqrt{v(1-v)} } \\ 
     & \approx & \ln(\OR) \sqrt{v(1-v)}\sqrt{w(1-w)}, \quad \,\,\,\, (\text{because} \,\,\, p-q \approx \ln(\OR)(1-v)v), \\ 
     & \approx & \delta.
\end{eqnarray*}
Standardized coefficients may be used in practice as a ``scale-free'' measure, however it has been suggested that their magnitude may not appropriately reflect relative importance of explanatory variables \cite{greenland1986fallacy, greenland1991standardized}. Since our $\gamma'$ is not obtained using a regular standardization technique (i.e., scaling by the standard deviation), we omit arguments for and against the use of standardized coefficients in statistical practice.

As we briefly discussed in the Methods Section, another two popular OR transformations with (-1, 1) range are Yule's $\mathcal{Y}$ and $\mathcal{Q}$ coefficients. If one were to apply the $\arctanh$ transformation to either $\mathcal{Y}$ or $\mathcal{Q}$, the result would be equal to $\log(\OR)$ times a constant. Thus, it should be expected that statistics that are directly based on $\mathcal{Y}$ or $\mathcal{Q}$, (without the variance stabilizing transformation), would not be competitive to the one based on $\log(\OR)$ -- the result that we confirmed via simulations. Curiously, although $\gamma'$ can be expressed as a function of $\mathcal{Y}$ (see Eq (\ref{eq:gamma_y})), statistical power of the $\gamma'$-test is higher than that of the $\log(\OR)$-based test, in stark contrast to properties of test statistics directly based on sample variances of Yule's coefficients. We further note that the relationship between Yule's coefficients and $\log(\OR)$ is monotonic for all values of $\log(\OR)$, while the relationship between $\gamma'$ and $\log(\OR)$ is monotonic only for log(OR) values varying between about $\nicefrac{1}{121}$ and $121$. Nonetheless, although the range of $\log(\OR)$ values admissible for $\gamma'$ transformation is limited to this range, it covers the majority of plausible $\log(\OR)$ values observed in practice.

We showed here how the standardized logarithm of odds ratio, $\delta$, can be utilized as a middle step towards an approximate posterior inference for a raw (non-standardized) parameter of interest. Assuming that the prior distribution for the raw parameter is known precisely, we checked performance of our method in terms of its resistance to the winner's curse and robustness of estimation in the presence of multiple testing. Of course, exact knowledge of the prior distribution is improbable, however it is a very useful assumption to make for the purpose of checking the accuracy of methods performance in such an ideal scenario. Assuming that the prior is known, proper posterior estimates should not overstate the effect size when the top-ranking associations are selected out of a large number of results. As for practical implementations, although the exact prior distribution may not be known, it is possible for it to be specified realistically. We recognize that the problem of a reasonable prior choice may be challenging, but also note that this problem is not unique to our method and is ubiquitous within the Bayesian framework. Therefore, in practice, it is important to assess robustness of posterior estimates to changes in prior parameters. For instance, in our application, we varied the prior probability of encountering a true association and found that the association between whole grain consumption and T2D was quite resilient to the increase in the prior $\Pr(H_0)$. The association between alcohol, fruits and berries, dietary magnesium or calcium consumption and T2D, on the contrary, vanished even under optimistic prior assumptions (75\% chance) about the frequency of real effects. 

Finally, we note that using the proposed methodology, the posterior estimates can be calculated using only commonly reported statistics (such as $\log(\OR)$ and its standard error), without requiring access to individual records. Within the proposed framework, any arbitrary prior distribution for a parameter that measures effect size can be easily accommodated in a form of a binned frequency table. The discretized nature of the prior does not preclude usage of continuous distributions, because modern statistical packages have facilities to finely chop continuous density functions, thus enabling one to obtain arbitrarily accurate approximations to continuous priors.

\section*{Acknowledgments}
Author affiliation: Biostatistics Department, University of Kentucky, Lexington, Kentucky, USA (Olga A. Vsevolozhskaya) \\
Correspondence: Dmitri V. Zaykin, Senior Investigator at the Biostatistics and Computational Biology Branch, National Institute of Environmental Health Sciences, National Institutes of Health, P.O. Box 12233, Research Triangle Park, NC 27709, USA. Tel.: +1 (919) 541-0096 ; Fax: +1  (919) 541-4311. Email address: dmitri.zaykin@nih.gov

This research was supported in part by the Intramural Research Program of the NIH, National Institute of Environmental Health Sciences, NIEHS. The authors would like to thank Gabriel Ruiz for his contribution to finding the LLC bound\cite{vrz2017bayesian} during a Summer of Discovery Internship at NIEHS.

\bibliography{Kepler}
\clearpage
\section*{Figure Legends}

\clearpage
\section*{Tables}
\begin{table}[th!]
  \centering
  \begin{tabular}[th!]{lcc} \hline
     & $\log(\OR)$ & $\gamma'$ \\ \cline{2-3}
    $n_D=25$   & 0.0290 & 0.0508 \\ 
    $n_D=50$   & 0.0381 & 0.0501 \\
    $n_D=100$  & 0.0432 & 0.0494 \\
    $n_D=250$  & 0.0464 & 0.0491 \\
    $n_D=500$  & 0.0476 & 0.0490 \\
    $n_D=1000$ & 0.0485 & 0.0490 \\  
    $n_D=5000$ & 0.0497 & 0.0498 \\ \hline
  \end{tabular}
  \caption{The Type-I error rate by the number of cases ($\log \OR = 0$).}
  \label{type1}
\end{table}

\begin{table}[th!]
  \centering
  \begin{tabular}[th!]{lcccccccc} \\ \hline
    & $\log(\OR)$ & $\gamma'$ & $\log(\OR)$ & $\gamma'$ & $\log(\OR)$ & $\gamma'$ & $\log(\OR)$ & $\gamma'$ \\ \hline
    $\log(\OR)$$\sim$$N(0,\tau)$ $\Rightarrow$ & \multicolumn{2}{c}{$\tau=\frac{\log(2)}{\Phi^{-1}(1-0.05)} \approx$  0.42} & \multicolumn{2}{c}{$\tau$ = 0.5} & \multicolumn{2}{c}{$\tau$ = 1} & \multicolumn{2}{c}{$\tau$ = 2} \\ \cline{1-9}
    $n_D=25$    & 0.065 & 0.098 &  0.080 & 0.116 & 0.212 & 0.263  & 0.441 & 0.493 \\
    $n_D=50$    & 0.121 & 0.142 &  0.151 & 0.174 & 0.358 & 0.385  & 0.602 & 0.624 \\
    $n_D=100$   & 0.204 & 0.217 &  0.253 & 0.266 & 0.503 & 0.516  & 0.718 & 0.726 \\
    $n_D=250$   & 0.360 & 0.365 &  0.423 & 0.429 & 0.664 & 0.668  & 0.821 & 0.823 \\
    $n_D=500$   & 0.490 & 0.493 &  0.553 & 0.556 & 0.757 & 0.758  & 0.873 & 0.874 \\
    $n_D=1000$  & 0.613 & 0.614 &  0.666 & 0.667 & 0.825 & 0.826  & 0.909 & 0.910 \\
    $n_D=5000$  & 0.814 & 0.814 &  0.843 & 0.843 & 0.921 & 0.921  & 0.960 & 0.960 \\ 
\hline
    Fixed OR $\Rightarrow$ & \multicolumn{2}{c}{OR = 1.25} & \multicolumn{2}{c}{OR = 2} & \multicolumn{2}{c}{OR = 3} & \multicolumn{2}{c}{OR = 4} \\ \cline{1-9}
    $n_D=25$   & 0.038 & 0.064 & 0.124 & 0.176 & 0.276 & 0.354 & 0.411 & 0.499 \\
    $n_D=50$   & 0.061 & 0.076 & 0.267 & 0.303 & 0.559 & 0.602 & 0.735 & 0.770 \\
    $n_D=100$  & 0.091 & 0.101 & 0.499 & 0.521 & 0.840 & 0.854 & 0.938 & 0.945 \\
    $n_D=250$  & 0.174 & 0.180 & 0.850 & 0.856 & 0.985 & 0.986 & 0.999 & 0.999 \\
    $n_D=500$  & 0.306 & 0.310 & 0.971 & 0.972 & 0.999 & 0.999 & 1     & 1     \\
    $n_D=1000$ & 0.532 & 0.534 & 0.998 & 0.998 & 1     & 1     & 1     & 1     \\ 
    $n_D=5000$ & 0.971 & 0.971 & 1 & 1 & 1     & 1     & 1     & 1     \\ 
\hline
  \end{tabular}
  \caption{Power of the tests by different levels of $\log(\OR)$ and $\tau$, assuming that $\log(\OR)\Nb \sim\Nb N(0,\tau)$.}
  \label{pwr}
\end{table}

\begin{table}[!th]
  \centering
  \begin{tabular}{llcccc}
    \hline
    \# of tests ($L$) & $n$ & $True E(\gamma')$ & $E(\gamma' \mid  Z_{\max})$ & $E(\widehat{\gamma'})$ & Posterior coverage \\ \hline
    10,000   & 500   &  0.46 & 0.46 & 0.55 & 94\% \\
             & 750   &  0.47 & 0.47 & 0.53 & 95\% \\
             & 1,000 &  0.48 & 0.48 & 0.52 & 95\% \\
             & 1,500 &  0.48 & 0.48 & 0.51 & 95\% \\ \hline
   100,000   & 500   &  0.52 & 0.53 & 0.62 & 93\% \\ 
             & 750   &  0.53 & 0.54 & 0.60 & 95\% \\
             & 1,000 &  0.54 & 0.55 & 0.60 & 95\% \\
             & 1,500 &  0.55 & 0.55 & 0.58 & 95\% \\ \hline 
   500,000   & 500   &  0.56 & 0.57 & 0.67 & 92\% \\
             & 750   &  0.56 & 0.57 & 0.64 & 94\% \\
             & 1,000 &  0.58 & 0.58 & 0.63 & 94\% \\
             & 1,500 &  0.59 & 0.59 & 0.63 & 95\% \\ \hline
  1,000,000   & 500  &  0.58 & 0.59 & 0.69 & 91\% \\
             & 750   &  0.58 & 0.59 & 0.66 & 93\% \\
             & 1,000 &  0.61 & 0.61 & 0.66 & 95\% \\ 
             & 1,500 &  0.61 & 0.61 & 0.65 & 95\% \\ \hline
  \end{tabular}
  \caption{Average true value, $E(\gamma')$, average posterior estimator, $E(\gamma' \mid  Z_{\max})$, and average frequentist estimator, $E(\widehat{\gamma'})$, assuming $\gamma'\sim N(0, \tau = 0.42)$ for the top-ranking (maximum) observed statistic ($Z$) selected out of $L$ tests. Averages refer to the mean value taken across simulation experiments}
  \label{tab:maxZ}
\end{table}

\begin{table}[th!]
\begin{adjustwidth}{-2cm}{}
 \begin{small}
   \begin{tabular}{lllll} \hline
    Dietary risk factor & OR/$\gamma'$, 95\%CI for $\gamma'$& \multicolumn{3}{c}{Posterior expectation and interval for $\gamma'$} \\ \hline
      & & $\pi_0 = 0.25$ & $\pi_0 = 0.5$ & $\pi_0 = 0.75$ \\ \cline{3-5}
     Whole grain intake \cite{fung2002whole}       & 0.70/-0.13, (-0.21, -0.06) & -0.13 (-0.20, -0.05)  & -0.13 (-0.20, -0.05) & -0.12 (-0.21, -0.03)  \\
     Protein intake\cite{tinker2011biomarker}      & 1.16/0.06,  (0.02, 0.09)   & 0.053 (0.01, 0.10)    & 0.05 (-0.00, 0.01)   & 0.04 (-0.00, 0.08) \\
     Alcohol consumption\cite{cullmann2012alcohol} & 1.67/0.19,  (0.04, 0.34)   & 0.15 (-0.00, 0.29)     & 0.13 (-0.00, 0.27)   & 0.10 (-0.05, 0.26)  \\   
     Fruits and berries\cite{montonen2005food}     & 0.69/-0.14, (-0.25, -0.03) & -0.12 (-0.23, 0.00)   & -0.10 (-0.21, 0.00)  & -0.08 (-0.20, 0.04)  \\
     Dietary magnesium\cite{hruby2014higher}       & 0.49/-0.26, (-0.47, -0.03) & -0.17 (-0.34, 0.01)   & -0.15 (-0.32, 0.03)  & -0.11 (-0.30, 0.09)   \\
     Dietary calcium\cite{van2006dietary}          & 0.86/-0.06, (-0.11, 0.00)  & -0.043 (-0.10, 0.01)  & -0.03 (-0.09, 0.03)  & -0.01 (-0.08, 0.05)  \\
  \end{tabular}
 \end{small}
  \caption{Posterior inference and robustness of results to prior assumptions: analysis of dietary risk factors associated with type-II diabetes.}
\end{adjustwidth}
\label{tab:res}
\end{table}
\clearpage
\appendix
\doublespacing
\section*{Supplemental material}
\renewcommand{\thesubsection}{S-\arabic{subsection}}
\setcounter{equation}{0}
\renewcommand{\theequation}{S-\arabic{equation}}

\subsection{Connection to the Laplace Limit Constant (Continued)}
To relate LLC to the bounds for standardized $\log(\OR)$, let $\psi = \log(\OR)$, $\psi\Nb >\Nb 0$. In terms of $\psi$, the maximum of the standardized statistic is given by
\begin{eqnarray}
  \gamma = \kappa(\psi) = \psi \left\{2 \sqrt{2 + \frac{1 +
  \exp{(\psi)}}{\exp{\left(\nicefrac{\psi}{2}\right)}}}\right\}^{-1} .
\end{eqnarray}
Using basic relations for hyperbolic functions:
\begin{eqnarray*}
   && \frac{1 + \exp{(\psi)}}{\exp{\left(\nicefrac{\psi}{2}\right)}} = 2 \cosh \left(\nicefrac{\psi}{2}\right) , \\
   && 2 + \frac{1 + \exp{(\psi)}}{\exp{ \left(\nicefrac{\psi}{2}\right) }} = 4 \left[ \cosh \left(\nicefrac{\psi}{4}\right) \right]^{2}, \quad \text{and} \\
   && \sqrt{2 + \frac{1 + \exp{(\psi)}}{\exp{ \left(\nicefrac{\psi}{2}\right) }}} 
      = 2 \cosh \left(\nicefrac{\psi}{4}\right) , 
\end{eqnarray*}
we can express $\kappa(\psi)$ and its first derivative as
\begin{eqnarray}
   \kappa (\psi) &=& \left(\nicefrac{\psi}{4}\right) \left\{\cosh \left(\nicefrac{\psi}{4}\right)\right\}^{-1} 
          = \left(\nicefrac{\psi}{4}\right) \, \sech \left(\nicefrac{\psi}{4}\right) \label{eq.kappa} \\ 
   \kappa^\prime (\psi) &=& \frac{4 - \psi \tanh  \left(\nicefrac{\psi}{4}\right) }
         {16 \cosh \left(\nicefrac{\psi}{4}\right) }.
\end{eqnarray}
To maximize the standardized log(OR), we need to set $\kappa^\prime(\psi)=0$, which is equivalent to solving 
$\left(\nicefrac{\psi}{4}\right)\Nb \tanh\Nb \left(\nicefrac{\psi}{4}\right)\Nb =\Nb1$ for $\psi$. The solution is four times the solution to $\psi\Nb \tanh(\psi)\Nb =\Nb 1$ equation, which is 1.19967864... This implies the maximum log(OR) $= 4 \times 1.1996...=4.7987...$, and by substituting this value into Eq. (\ref{max.gamma}) we obtain $\gamma_{ \max } = 0.6627...$, the LLC. The solution to KE 
involves the condition equivalent to Eq. (\ref{max.gamma}). Namely, the solution can be expressed as the power series in $\varepsilon$, provided $|\varepsilon \sin(E)| < |E-M|$ and that $\varepsilon < \psi / \cosh(\psi), \psi = |E-M|$, which is the LLC \cite{plummer1918introductory}.

\subsection{Simulation Setup} \label{simulations}
We used simulations to compare performance of the asymptotic test based on the $Z=\frac{\log(\widehat{\OR}) }{ \sqrt{ Var (\log(\widehat{\OR}))}}$ statistic to the one based on the new statistic, $\gamma'$. For simulation $i = 1, \ldots, 10^6$, the true $\log(\OR)$  was assumed to be either (a) fixed and equal to the same value across simulations, or (b) normally distributed around zero with the standard deviation $\tau$. The parameter $\tau$ was chosen as $\tau=\log(2)/\Phi^{-1}(1-0.05) \approx 0.42$, that is, we assumed that there is 5\% {\em a priori} chance of encountering OR $\ge2$, and 5\% chance of encountering OR $\le \nicefrac{1}{2}$.

The number of cases $n_D$ was (25, 50, 100, 250, 500, 1000, 5000) and the number of controls was generated as $n_{\bar{D}} \sim \left[\text{Unif}( \nicefrac{1}{2} n_D, n_D) \right]$, where $\left[\cdot\right]$ is the nearest integer function. The probability of exposure among cases was drawn from a uniform distribution, $p \sim \text{Unif}(0.05, 0.95)$, and the corresponding probability of expose among controls was calculated as $q = p/\left[ (1-p)\OR + p\right]$. Two-by-two table cell counts were obtained under the binomal sampling with $ \nicefrac{1}{2}$ added to each cell count:
\begin{eqnarray*}
  n_{11} &\sim&  \nicefrac{1}{2} + \text{Binomial}(n_D, p), \\
  n_{21} &\sim&  \nicefrac{1}{2} + \text{Binomial}(n_{\bar{D}}, q), \\
  n_{12} &=& \nicefrac{1}{2} + n_D - n_{11}, \\
  n_{22} &=&  \nicefrac{1}{2} + n_{\bar{D}} - n_{21}.
\end{eqnarray*}
The addition of $\nicefrac{1}{2}$ to cell counts is known as the Haldane-Anscombe correction, commonly used to improve asymptotic convergence to normality of the test statistic for log(OR) \cite{haldane1956estimation,anscombe1956estimating,lawson2004small,agresti1999logit}. It can also be shown that after this correction, the resulting variance estimator, $\Var\Nb \left[\Nb {\log\Nb (\Nb\widehat \OR\Nb )}\Nb \right]$, approximates the posterior variance estimator derived assuming Jeffreys' prior, i.e., Beta($ \nicefrac{1}{2}$,$ \nicefrac{1}{2}$) prior distribution for $p$ and $q$ \cite{zaykin2004interval}.

For studying Bayesian properties, simulated data sets were obtained as described above, but with the following modifications. We sampled $\psi\Nb =\Nb \log(\OR)$ from a prior mixture distribution, where $\psi$ was equal to zero with probability $\pi_0\Nb =\Nb\Pr(H_0)\Nb=0.8$, and with probability $(1-\pi_0)$, the values $\psi$ were sampled as $\psi\Nb \sim\Nb \text{Truncated\Nb Normal}\Nb(\Nb 0\Nb,\Nb t)$, discretized into 100 bins. The truncation parameter was set to $t$=4.8, which corresponds to the maximum OR of about 121. In each simulation run (out of 10,000 in total for each setting), we performed $L = 10^4, 10^5, 5\times 10^5$, or $10^6$ tests and calculated posterior estimates for the top-ranking result based on the largest value of a $Z$-statistic out of $L$ tests.
\end{document}